\documentclass[twocolumn,showpacs,preprintnumbers,amsmath,amssymb]{revtex4}

\usepackage{graphicx}
\usepackage{dcolumn}
\usepackage{bm}

\begin{document}

\title{Addressing geometric non-linearities with cantilever MEMS: beyond the Duffing model}

\author{E. Collin} 
\email{eddy.collin@grenoble.cnrs.fr}
\author{Yu. M. Bunkov}
\author{H. Godfrin}

\affiliation{%
Institut N\'eel
\\
CNRS et Universit\'e Joseph Fourier, \\
BP 166, 38042 Grenoble Cedex 9, France \\
}%

\date{\today}

\begin{abstract}
We report on low temperature measurements performed on micro-electro-mechanical systems (MEMS) driven deeply into the non-linear regime. 
The materials are kept in their elastic domain, while the observed non-linearity is purely of geometrical origin.
Two techniques are used, harmonic drive and free decay. For each case, we present {\it an analytic} theory fitting the data. The harmonic drive is fit with a {\it Lorentz-like} lineshape obtained from an extended version of Landau and Lifshitz's non-linear theory. The evolution in the time domain is fit with an {\it amplitude-dependent frequency} decaying function derived from the Lindstedt-Poincar\'e theory of non-linear differential equations. 
The technique is perfectly generic and can be straightforwardly adapted to {\it any} mechanical device made of ideally elastic constituents, and which can be reduced to a single degree of freedom, for an {\it experimental} definition of its non-linear dynamics equation.
\end{abstract}

\maketitle

\section{INTRODUCTION}

Micro and nowadays nano-mechanical systems (MEMS and NEMS) are of current interest due to their broad field of scientific and technical applications.
 {\it All devices} (and evidently not only mechanical ones) present a non-linear behavior at large drives, and a large panel of scientists from different communities is dealing today with non-linear mechanics \cite{lifshitzcross,roads}. 
 
From an engineer's point of view, non-linearity is a key design parameter. When used in the linear regime, non-linearity limits the dynamic range of a device \cite{roukesnonlin}. One can also exploit non-linearity with for instance frequency mixing \cite{mix}, synchronization \cite{sync}, amplification using bifurcation points \cite{ampli}, suppression of amplifier noise in oscillator circuits \cite{oscill, oscillnoise, noise2}, and mass (homodyne) detection \cite{mass1}. 
Moreover, the non-linear component proves to be essential to complex, useful and efficient designs, with for instance the diode in conventional electronics and the Josephson junction in superconducting circuitry \cite{michel}. 

From a physicist's point of view, a MEMS/NEMS with a canonical non-linearity is a close realization of the {\it Duffing oscillator}, a mechanical system with a spring force containing a term $F_{non-lin.} \propto x^3$ ($x$ displacement). It is of ubiquitous interest in physics since many systems can be mapped on this problem \cite{Gabrielse, irfan}. It also provides a simple mathematical model which is in many cases {\it analytically} solvable.
Furthermore, the simple Duffing expression enables the theoretic and, with the MEMS/NEMS close implementation, the {\it model experimental} study of subtle dynamic properties like dynamical switching \cite{switch1,switch2, roukes_prl} and memory effects \cite{memory}. 

Due to the fundamental issue behind non-linear dynamics (the physics of chaos \cite{lifshitzcross}) and the broad panel of applications in micro/nano mechanics, which can even be extended to the {\it quantum-limited} nano-mechanical device \cite{lifsh_quant}, it is important to understand the {\it nature} of these mechanical non-linearities \cite{roukes_prl}. The most commonly discussed cases are non-linear actuation with an electrostatic drive \cite{nonlin_electrostat, superharm}, and non-linear constituents (with i.e. an ${x}^2$ term in the damping \cite{noise2,lifshitzcross,sync}). Clever designs making use of these non-linearities enable parametric amplification \cite{prl_parametric, mass2}, and parametric drive \cite{parametricdrive}. In particular, non-linear dampings \cite{jeevak_opto} permit the realization of a mechanical {\it Van der Pol} oscillator \cite{vanderpol}.

Furthermore, even with perfectly {\it elastic} constituents (Young modulus and damping independent of strain/stress), mechanical devices do present a non-linear behavior which is effectively captured by the Duffing model. This non-linearity is of pure {\it geometrical} origin \cite{book1,book2}, and in its most general form it will contain other terms in the dynamics equation in addition to the cubic restoring force, with a straightforward second-order force $F_{non-lin.} \propto x^2$ \cite{landaumeca, geom}, and less intuitively {\it inertia} non-linear terms \cite{crespo,clelandbook,geom2}. To put it in crude words, the Duffing equation is, even for these perfectly elastic devices, only a convenient model describing correctly the measurements. In practice a mechanical device {\it is certainly not} a Duffing oscillator.

However in practice, most of the theoretical and experimental work has been done around the Duffing problem (e.g. \cite{roukesnonlin,mix,ampli,oscill,oscillnoise,mass1,roukesAPL_duffing}), with most of the experiments done in the {\it driven} regime (e.g. \cite{roukesnonlin,roukesAPL_duffing,geom,coating,JLTP_VIW}).
The reason behind this fact is certainly simplicity. Theoretically, the full non-linear problem is extremely complex while a simple Duffing modeling does capture the observed mechanical behavior. Experimentally, the driven regime is easier to handle since it uses the natural amplification of the system through its $Q$ factor.
\begin{figure}[t!]
\includegraphics[height=5.5 cm]{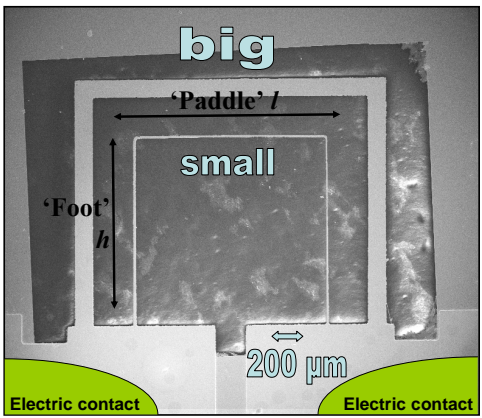}
\caption{\label{pictureSEM} (Color online) 
SEM (Scanning Electron Microscope) picture of the 'double-structure' sample studied in the present article. The larger goal-post oscillator is approximately 8.5 times wider than the smaller (width $\approx 15~\mu$m), both having dimensions $l \approx h \approx 1.5~$mm. The average silicon thickness of the structures is about $6.5~\mu$m, with 200$~$nm metal on top.}
\end{figure}
%
%

The most sophisticated analytical modelings derive from basic continuum mechanics a system of coupled non-linear equations describing the generic dynamics of the mechanical device \cite{book2}. 
This system is usually reduced to a one dimensional problem by a Gal\'erkin-type procedure or a normal mode expansion of the linearized problem \cite{book2,geom2}. For beams, most of the recent work is based on the formulation of Crespo da Silva and Glynn \cite{crespo}. In the case of thin and long beams and planar motion, extended non-linear Euler-Bernouilli equations are available \cite{book1}.

The classical procedure is to compute the response of a system from Crespo da Silva-like equations, and since even the reduced analytic writing is very complex the final step is performed numerically \cite{ahmadian, geom2, superharm, nonlin_electrostat, gottlieb}. In these works, the aim is to {\it predict the dynamic behavior} of a particular device from its mechanical characteristics, the final product being a numerical curve plotted on experimental points proving good agreement. Here, we adopt a radically different approach inspired from low temperature physics. Our aim is to predict {\it the analytical shape of the most generic dynamics equation} for an ideally elastic non-linear mechanical device that has been reduced to a single degree of freedom. This equation shall contain a number of non-linear coefficients, each of which having a well defined meaning. But we shall {\it not} compute these coefficients, and leave them as characteristics of the devices that have to be obtained by other means. We present the exact analytical solutions of the full dynamics equation in two relevant experimental cases: harmonic drive and free decay. 
We demonstrate on simple cantilever MEMS devices that these expressions can be used to {\it fit experimental data} and extract non-linear parameters.
The strength of our new approach lies in {\it its completely generic nature, and purely analytic formulation}. 

\section{EXPERIMENTAL RESULTS} 
\label{exp}

We present measurements on perfectly elastic cantilever-based MEMS devices extending deeply in the non-linear regime of their first resonant mode. Two techniques are used: a frequency-sweep technique where the device is continuously driven with a harmonic force $F(t)=F_0 \cos(\omega t)$, and a time-decay method where the oscillator is suddenly released with an initial displacement/velocity (signal recorded  under $F(t)=0$). Note that beyond the linear regime, the two measurements {\it are not} the Fourier transform of each other anymore.

The sample studied in the present article is the 'double-structure' of Ref. \cite{coating}, shown in Fig. \ref{pictureSEM}. It consists of two micro-mechanical goal-post silicon structures \cite{JLTP_VIW} (hereafter called 'big' and 'small') embedded one in the other. Each structure is made of two cantilevers ('feet' of length $h$) linked by a 'paddle' of length $l$.
The measurements are performed using the magnetomotive scheme. 
The low temperature condition (4.2$~$K) is a practical means to use moderately high magnetic fields, cryogenic vacuum ($< 10^{-6}~$mbar), and obtain low electrical noise. 
A static magnetic field $B$ is imposed along the sample while a current $I(t)$ is fed through the thin (non-superconducting) metallic layer that covers it. A time-dependent Laplace force of amplitude $F(t)=I(t) l B$ acts on each structure driving it out of the plane. 
Their motion induces in turn a voltage 
$V(t)=l B v(t)$ proportional to the velocity $v(t)=\dot{x}(t)$ of the 'paddles' (Lenz's law).
Note that for our devices, the driving force $F$ is ideally linear, as opposed to electrostatic actuation \cite{nonlin_electrostat, superharm, geom2}. The detected signal however is weakly non-linear at large bendings, but this effect can be proven to be negligible \cite{JLTP_VIW}.

The double-structure design enables the simultaneous study of two very different oscillators, with very different resonance frequencies. 
No relevant mechanical coupling between the two could be detected, even deeply in the non-linear regime, meaning that the resonance peaks are perfectly well separated. 
However, the structures are electrically coupled because of their parallel wiring. The drive current splits in two according to the electric resistance $R_b$,$R_s$ of each structure, while the detected voltage is reduced by the same proportion (Kirchhoff's rules). The ratio of these resistances measured experimentally follows accurately the geometrical dimensions of the structures.
Moreover, the induced voltage that we detect (proportional to $B^2$) generates in turn a loop current through the total resistance $R_b+R_s$. This effect (which opposes the driving force) has to be taken into account for large magnetic fields.
Careful calibration of the setup enables the definition of {\it absolute} displacements $x$ of the 'paddle' in $\mu$m, and applied forces $F$ in pN, quoted here in {\it peak values}.
Experimental details can be found in \cite{JLTP_VIW,coating}.

\subsection{Generic description}

\begin{figure}[h!]
\includegraphics[height=2.5 cm]{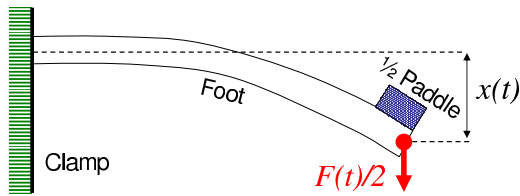}
\caption{\label{picture} (Color online) 
{\it Equivalent} loaded cantilever (one 'foot' of the goal-post structure), with half the 'paddle' mass as end load (normal mode mass $m/2$, normal spring constant $k/2$ and dissipation constant $\Lambda/2$). The parametrization uses the end displacement $x(t)$ with $v(t)=\dot{x}(t)$.}
\end{figure}

\begin{figure}[!h]
\includegraphics[height=6. cm]{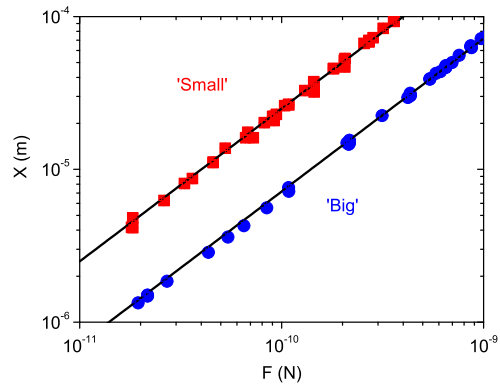}
\caption{\label{elast} (Color online) 
Maximal displacements versus force measured with the harmonic drive method, at 4.2$~$K in vacuum. Circles (blue) 'big' structure, and squares (red) 'small' structure, for various currents and fields. The data correspond to the height of the resonance peak obtained for upwards frequency sweeps (see text). The straight lines prove that the relation $x_0=F_0/k \, Q$ holds (Section \ref{theory}), with a constant quality factor $Q$ (i.e. damping) \cite{coating, JLTP_VIW}.}
\end{figure}

\begin{figure}[!h]
\includegraphics[height=5. cm]{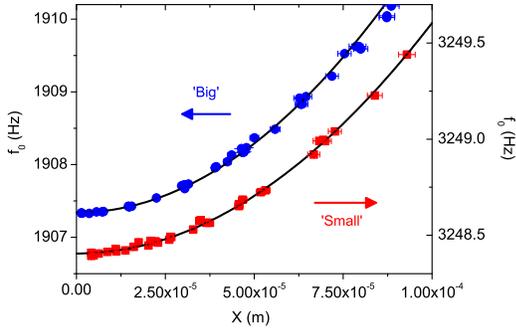}
\caption{\label{quadfreq} (Color online) 
Resonance frequencies versus displacement measured for the two oscillators using the harmonic drive technique, in vacuum at 4.2$~$K. Circles (blue) stand for the 'big' structure, and the squares (red) for the 'small' one. The data are taken while sweeping the frequency upwards. The black lines are quadratic fits (Section \ref{theory}) yielding the non-linear pulling parameters $\beta$ \cite{JLTP_VIW}. Note the different axes.}
\end{figure}

In the simplest analytical approach, one can demonstrate that each oscillator is almost {\it equivalent} to a mass-loaded cantilever (Fig. \ref{picture}) \cite{JLTP_VIW}. 
%
%
We keep the displacements $x$ in the linear regime of the constitutive materials of the cantilevers (silicon and coating). 
For the present work dealing with {\it mm} long structures, it corresponds to about $x \leq 100~\mu$m ({\it peak values}).
This is easily verified experimentally by measuring the maximal displacement in the frequency domain as a function of the force, and verifying that the damping remains independent of the strain \cite{coating}, Fig. \ref{elast}. At the same time, no anomalous frequency shifts (over the the geometrical term described below, Fig. \ref{quadfreq}) could be detected.

However, as any other mechanical structure (let it be e.g. a torsional rod, a doubly-clamped beam, or an STM tip), the dynamics of each oscillator around its first mechanical mode follows a non-linear equation, which {\it most generic} expression for geometrical non-linearities, expanded at 3$^{rd}$ order, is given in Section \ref{theory}, Eq. (\ref{nonlin}), from absolutely basic considerations. In the simplified version relevant to our experiments (see Appendix), it writes:%

\begin{widetext}
\vspace*{-7 mm}
\begin{equation}
 m\left(1+ m_1 \, x + m_2 \, x^2 \right) \, \ddot{x}  +  m \left(\frac{1}{2} m_1+ m_2 \, x \right)\dot{x} ^2 +
 2 \Lambda \, \dot{x}  +  k \left( 1+ k_1 \, x + k_2 \, x^2\right) \,  x  =F(t),  \label{dyneq}
\end{equation}
\end{widetext}

%

$x$ being the displacement of the top part of the structures (the 'paddle').
$2 \Lambda\, \dot{x}$ is the damping term arising from the friction mechanisms present in the devices, with $m$ and $k$ the normal mass and spring constant of the mode under study.
Since the materials are in their linear regime,
$m$, $k$ and $\Lambda$ are drive/displacement independent; the only non-linear terms in the equation are the $m_i$ (inertial) and $k_i$ (elastic) constants. 

%
%

The coefficients $m_i$ and $k_i$ are specific to the exact nature of the device under study (see Section \ref{theory} and Appendix).
These terms appear naturally for large cantilever distortions, as {\it characteristics of the geometrical non-linearity} \cite{book1,book2,clelandbook}. 
We confirmed this origin for our experiments: the non-linear signatures are temperature-independent \cite{JLTP_VIW}, and for samples having the same aspect ratios (while having different damping, resonance frequency, and metallic coating) the non-linear coefficients follow the same geometrical scalings (see discussion in Section \ref{comps}).

Our theoretical modeling considers {\it the full 1D non-linear expression of the dynamics} (\ref{dyneq}), with an {\it inertia} non-linearity and a non-linear {\it restoring force}, all up to order 3 in the displacement. 
We give in the following theoretical tools enabling the fit of the data and the determination of non-linear parameters for two experimental cases: harmonic drive and free decay (Section \ref{theory}). 
Free decay and harmonic drive measurements have been used by Gottlieb {\it et al.} \cite{gottlieb} in order to carefully characterize the drag force of air on a cantilever STM. A full non-linear model based on Crespo da Silva \cite{crespo} was used and solved numerically, including in addition non-linear damping.\\
To our knowledge, our work is the first one presenting an {\it analytic full solution to the generic Eq. (\ref{nonlin2}), providing tools to extract experimentally intrinsic information on geometrical non-linearities}.

\subsection{{\it Case 1} - Harmonic drive $F(t)=F_0 \cos(\omega t)$}
We extend Landau \& Lifshitz's \cite{landaumeca} non-linear approach, revisiting the results of \cite{JLTP_VIW}. 
We postulate for the solution
a superposition of oscillating terms $\cos (n\,\omega t + \phi_n)$, 
and seek only the first one $n = 1$. Replacing the above expression in Eq. (\ref{dyneq}), we obtain for the voltage response a {\it modified Lorentzian} lineshape:
\begin{eqnarray*}
X & = & l B \omega \frac{F_{0}}{k} \frac{\Delta \omega \, \omega/(\omega_0)^2}{\left[(\omega_r/\omega_{0})^2-(\omega/\omega_0)^2 \right]^2+ \left[\Delta \omega \, \omega/(\omega_0)^2\right]^2}, \label{nonlinquad} \\
Y & = & l B \omega \frac{F_{0}}{k} \frac{(\omega_r/\omega_{0})^2-(\omega/\omega_{0})^2}{\left[(\omega_r/\omega_{0})^2-(\omega/\omega_0)^2 \right]^2+ \left[\Delta \omega \, \omega/(\omega_0)^2\right]^2}.   \label{nonlinphase}
\end{eqnarray*}
$X$ and $Y$ correspond to the amplitude of the signal in-phase, and out-of-phase with the driving force respectively.
The (angular) mode resonance frequency is $\omega_0=\sqrt{k/m}$, and $\Delta \omega =2 \Lambda/m$ the full width at half height of the {\it linear} resonance line $X$ (obtained for small displacements). 

%
\begin{figure}[!t]
\includegraphics[height=6.6 cm]{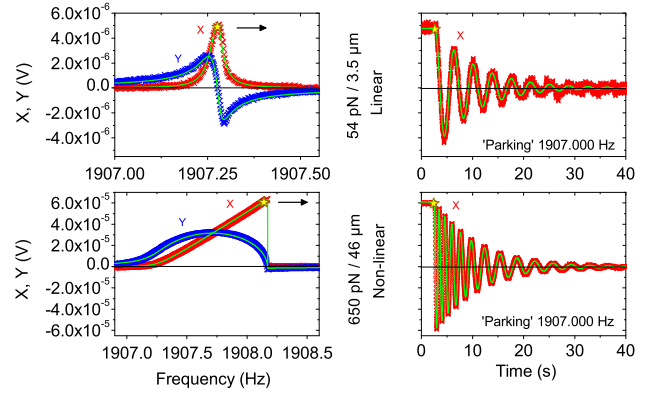}
\caption{\label{biglines} (Color online) 
Signals recorded at 4.2$~$K in vacuum on the first mechanical mode of the 'big' structure. The magnetic field used is 101$~$mT. $X$ denotes a signal in-phase, and $Y$ a signal out-of-phase with the driving force.
Left: harmonic drive method. Right: free decay method. The harmonic driving force is given in the graph with the deflection amplitude at maximum (in {\it peak values}). 
The star represents the point on the resonance where the oscillator was released for the corresponding free decay measurement.
Full (green) lines are fits explained in the text, giving $m =4.9~\mu$g, $f_0 = 1907.35~$Hz, $\Delta f =40~$mHz, $\tau = 8.0~$s, $\beta = +2.1\,10^5~$ m$^{-2}$ and $\lambda= +2.3\,10^5~$ m$^{-2}$. Error bars are typically $\pm 5~$\%, apart for the resonance frequency which is known within $\pm 10~$mHz.}
\end{figure}

The resonance position $\omega_r$ introduced in the above expressions is now a function of the amplitude of the displacement $x_0$. We write 
$\omega_r   =   \omega_0 (1+ \beta \, x_0^2)$ with $\beta$ the {\it frequency pulling term} :
\begin{eqnarray}
\beta &= & +\frac{3}{8} k_2 -\frac{1}{4} m_2 + \left(\frac{1}{12} k_1 -\frac{1}{8} m_1\right)\left(k_1-\frac{3}{2} m_1\right) \nonumber \\
&&-\left( \frac{1}{2} k_1 - \frac{1}{4} m_1  \right) \left(k_1-\frac{1}{2} m_1 \right) \label{beta}
\end{eqnarray}
written here for $x_0$ given in {\it peak values}. This expression is obtained in 
the underdamped regime ($Q=\omega_0/\Delta \omega$ $>\!\!> 1$) from the $\beta_1$ formula of Section \ref{theory}, Eq. (\ref{beta1}). In this high-$Q$ limit, when the amplitude of the displacement $x_0$ is small, the above $X$ and $Y$ peaked functions reduce to the simple {\it Lorentz} line. 
But when $x_0$ increases beyond a critical value $x_c$, the functions become bi-valued \cite{landaumeca}: two different branches are measured sweeping the frequency up, or down. The Lorentz line is distorted, pulled up or down depending on the sign of $\beta$ (Fig. \ref{quadfreq}) \cite{landaumeca}; besides, the height of the resonance peak $X$ measured while sweeping the frequency in the pulling direction (i.e. upwards sweep for $\beta>0$) remains proportional to $F$ and inversely proportional to the damping term $\Delta \omega$ \cite{JLTP_VIW,coating} (Fig. \ref{elast} and Section \ref{theory}).\\

In practice, the measurement technique is the well-known to low temperature physicists magnetomotive "vibrating wire" scheme \cite{ref_viw}. A current $I_0 \cos(\omega t)$ is fed through the structure. We monitor with a lock-in amplifier the voltage $X=l B v_0 \cos \phi$ in-phase with the excitation $F$, and the out-of-phase component $Y=l B v_0 \sin \phi$ (with $v_0$ the velocity amplitude of the 'paddle', and $\phi$ the phase). 
The displacement amplitude $x_0$ is obtained through $x_0=v_0/\omega_0$ to a very good accuracy.
The measurement is performed by sweeping the frequency $\omega$ {\it upwards} as slowly as possible, while recording $X,Y$. Two typical resonance lines for the 'big' and 'small' oscillators are shown in Fig. \ref{biglines} and \ref{smalllines} respectively.
The {\it modified Lorentzian} lineshape, solved and fit on the data (Section \ref{theory}), yields the mass $m$, spring constant $k$, linewidth $\Delta \omega$ and the non-linear coefficient $\beta$. This parameter fit on the line is the same as the one extracted from Fig. \ref{quadfreq}. Results are summarized in the captions of Figs. \ref{biglines},\ref{smalllines}. Details on the theoretical tools are given in Section \ref{theory}, and \cite{JLTP_VIW}.
Note the quality of the fits (backbone curves): the root-mean-square average error between data and fit ($\chi^2$) is typically smaller than a couple of \% of the maximal height.

\subsection{{\it Case 2} - Free decay $F(t)=0$}
We apply the Lindstedt-Poincar\'e method of solving non-linear differential equations \cite{lindstedt}. The novelty here is that we give the  solution to {\it the full equation Eq. (\ref{dyneq}), in the presence of a damping term}. 
We write for the solution $x(t)  =  s_0(t) \, + \delta s_{\lambda_i}(t)$, with $\delta s_{\lambda_i}(t)$ a perturbative function depending (at first order linearly) on non-linear coefficients $\lambda_i$ which are combinations of the $m_i,k_i$.
 The method is based on the idea that the function $s_0(t)$ should involve an oscillation frequency $\omega_R$ which is also a series of the $\lambda_i$, written in such a way that the expansion of the function {\it should cancel all secular terms} which would remain in a standard perturbative theory \cite{lindstedt}. We thus introduce 
$ \omega_R = \omega_0 + \delta  \omega_{\lambda_i} (t)$,
 with $\delta  \omega_{\lambda_i} (t)$ the unknown to be defined. By construction, the solution $s_0$ is:
\begin{displaymath}
s_0(t) = X_0 \exp \left(-  \frac{\omega_R}{\omega_0} \frac{t}{\tau} \right) \cos \left(\sqrt{1-\left(\frac{1}{\omega_0 \tau}\right)^2} \omega_R \, t + \varphi \right)
\end{displaymath}
with $\tau=2/\Delta \omega$ the ring-down time ($X_0$ and $\varphi$ are initial conditions). 
Canceling the secular terms brings finally $\omega_R = \omega_0 \, [1  + \lambda \, X_0^2 \, E(t)]$ with only one {\it pulling term} $\lambda$ (see Section \ref{theory} and Eq. (\ref{lambda1}), the $\lambda_1$ definition): 
\begin{eqnarray}
\lambda & = & +\frac{3}{8} \left(k_2 + m_1^2 - k_1 m_1 -m_2\right) + \frac{1}{8} \left(m_2-\frac{1}{2} m_1^2\right) \nonumber \\
&&   -\frac{5}{12} (k_1-m_1)^2 -\frac{1}{24} m_1^2 -\frac{5}{24} m_1 (k_1-m_1) , \label{lambda}
\end{eqnarray}
with the amplitude $X_0$ given in {\it peak values}, and $E(t)=\left[ 1 - \exp \left(- t/(\tau/2) \right)\right]/\left[t/(\tau/2) \right] $. 
The expression reduces to the linear result when $\lambda=0$ as it should. 
The function $\delta s_{\lambda_i}(t)$ brings oscillations at $2 \omega_R$, $3 \omega_R$ plus a 'constant', all decaying with exponential prefactors. 
\begin{figure}[!t]
\includegraphics[height=6.6 cm]{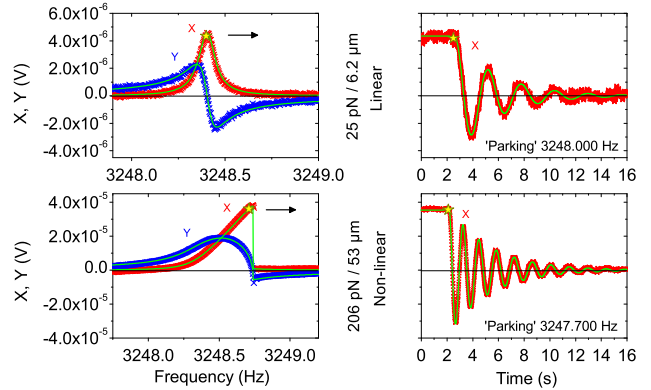}
\caption{\label{smalllines} (Color online) 
Same as Fig. \ref{biglines} but for the 'small' structure. The magnetic field is 351$~$mT. Fits (full/green lines) give $m=0.36~\mu$g, $f_0 = 3248.43~$ Hz, $\Delta f = 100$ mHz, $\tau = 3.2~$s, $\beta = +0.37\,10^5~$ m$^{-2}$ and $\lambda=+0.34\,10^5~$ m$^{-2}$.}
\end{figure}
To our knowledge, the mathematical treatment used to fit the data is original. Since the derivation of the formulas is non-trivial, we give the calculation details in Section \ref{theory}. The results are robust in the underdamped regime $Q=\omega_0 \tau / 2>\!\!>1$, and can be applied to {\it any weakly non-linear oscillator} described by Eq. (\ref{dyneq}).

In practice, we use a measurement scheme adapted from Pulsed-NMR \cite{ref_nmr}. The oscillators are put into motion by stopping a frequency-sweep close to the maximum of the resonance peak (stars in Figs. \ref{biglines} and \ref{smalllines}). The drive frequency is then suddenly switched to a lower ('parking') value $\omega_0 - \delta \omega$ where the mode is not excited. The ring-down signal is then recorded with a lock-in amplifier through the in-phase {\it beating} $X$ of the devices' oscillation with the drive ($\varphi=0$). The high filtering of the lock-in restitutes the $s_0(t)$ component alone, and provides an extremely large signal-to-noise ratio that cannot be attained in a direct ring-down acquisition \cite{gottlieb}. With $V(t) = l B \dot{s_0}(t)$, the expressions above are used to fit the data, adjusting the parameters $\omega_0$, $\tau$ and $\lambda$.
Results are summarized in Figs. \ref{biglines},\ref{smalllines} captions. The fits are very good (backbone curves), with an average root-mean-square error smaller than typically 4$~$\% of the maximal amplitude. \\

\subsection{Discussion}
\label{comps}

In the preceeding, we demonstrated that the theoretical tools of Section \ref{theory} can be used to fit experimental data. The dynamical parameters can be obtained by both a frequency sweep or a free decay technique 
(captions of Figs. \ref{biglines} and \ref{smalllines}), leading to the mode's mass $m$ and spring constant $k$ (through the height of the signal and the resonance frequency $\omega_0$), dissipation constant $\Lambda$ (through the linewidth $\Delta \omega$ or decay time $\tau$), and the non-linear pulling term (through $\lambda$ or $\beta$). 

We have studied many structures having almost the same aspect ratios, but with different metallic coatings and resonance frequencies: all 'small-like' samples display similar geometrical non-linear coefficients scaling as $\beta \, h^2 \approx + 0.065 ~ \pm 10 ~ $\% (for {\it peak values}) \cite{coating, JLTP_VIW}.
For the 'big' structure, the non-linear coefficients $\beta,\lambda$ are about 5.5 times larger.

\section{THEORETICAL TOOLS}
\label{theory}

We consider a mechanical structure that can be described by {\it standard continuum dynamics}.
We assume that the device is 
made of {\it ideally linear materials} (i.e. constant damping and elastic moduli with respect to the induced stress/strain). 
For large distortions, the structure is nonetheless non-linear for {\it geometrical} reasons \cite{JLTP_VIW,geom2,book1,book2,geom,clelandbook,crespo,ahmadian}. In order to keep the discussion as simple as possible, we restrict ourselves to the case of rectangular thin and long beams, neglecting the Poisson ratio. 
The problem reduces then to a single degree of freedom, and can be described by the tools of beam dynamics \cite{beamdyn}.
On the other hand, we discuss both the cantilever (inextensional) and the doubly clamped bridge (extensional), while the work of \cite{crespo} is restricted to inextensional beams only.
Note that the approach developed here can be straightforwardly adapted to other types of (simple) structures.

The idea behind our modeling is to consider the device's dynamically distorted shape, for mode $i$, as {\it a given}. 
A weaker form of the argument, based on the Rayleigh method, was presented in \cite{JLTP_VIW}.
We write it $f_i$, and since we will be interested only in the fundamental mode $i=0$, we drop immediately the index $i$.
Since the materials are perfectly {\it elastic}, we can write:
\begin{equation}
f\left(z,x(t)\right), \label{fund}
\end{equation}
the overall amplitude of the distortion being parametrized by $x(t)$ (and $z$ is the coordinate along the beam), a displacement which is typically the maximal deflection of the structure (i.e. the end tip of a cantilever, or the middle part of a doubly clamped beam in their first resonant mode).

We {\it shall not} be concerned with the computation of $f$. Suffice it to say that it is the solution of a continuum dynamics modeling based primarily on the extended Hamilton principle \cite{hamilton}, which can be of the type of \cite{crespo}, or of an extension of Euler-Bernoulli's beam equations \cite{book1,book2}. We present below the energetic writing resulting from this non-linear formulation.

The geometrical non-linearity arises straightforwardly from
the full expression of the distortion's radius of curvature $r^{-1}=\partial^2 f/\partial z^2 /[1+(\partial f/\partial z)^2]^{3/2}$, and from the elongation $dl/dz$ of the centroid:
\begin{equation}
\frac{d l (z,x(t))}{dz} = \sqrt{ 1+\left( \frac{\partial \, f\left(z,x(t)\right)}{\partial z} \right)^2 }. \label{dldz}
\end{equation}
The function $f$ is regular, and for a given parametrization $x(t)$ it is unique.
If the overall displacement $x(t)$ is not too large, we can take a Taylor series expansion in $x(t)$ of Eq. (\ref{fund})
, here at third order:
\begin{eqnarray*}
& & f\left(z,x(t)\right)  =  f(z,0) + \frac{\partial f(z,x=0)}{\partial x} x(t) \\
& & + \frac{1}{2}\frac{\partial^2 f(z,x=0)}{\partial x^2 } x(t)^2 + \frac{1}{6} \frac{\partial^3 f(z,x=0)}{\partial x^3} x(t)^3.
\end{eqnarray*}
$f(z,0)$ is the static distortion (we limit the discussion to $f(z,0)=0$, the straight beam) and $\partial f(z,x=0)/\partial x$ is the mode shape in the linear regime. The expressions below and Eq. (\ref{dldz}) will be developed in series on the same footing.

The length $h_l$ of the dynamically distorted beam is defined by:
\begin{displaymath}
h_l = \int_{0}^{h_z} \frac{d l(z,x(t))}{dz} \, dz.
\end{displaymath}
For a cantilever structure, the integration length $h_z$ has to be defined by the additional condition that the total length $h_l$ of the beam remains constant:
\begin{eqnarray*}
h_z &= & h \, (1 - h_2\, x(t)^2 - h_3\, x(t)^3) ,\\
h_l &=& h,
\end{eqnarray*}
while for a doubly clamped beam, the integration length is fixed and the centroid elongates:
\begin{eqnarray*}
h_z &= & h ,\\
h_l &=& h \, (1 + h_2\, x(t)^2+h_3\, x(t)^3),
\end{eqnarray*}
the development involving the same coefficients $h_i$ in the two cases (with obviously $h_2>0$).

The integrated kinetic energy $E_c(t)$ and potential energy $E_p(t)$ are:
\begin{eqnarray*}
E_c(t) & = & \frac{1}{2} \int_{0}^{h_z}  \left( \frac{\partial \, f\left(z,x(t)\right)}{\partial t} \right)^2 
( \rho \, w e ) \,\frac{d l(z,x(t))}{dz} \, dz , \\
E_p(t) & = & \frac{1}{2} \int_{0}^{h_z}  \left( \frac{\partial^2 \, f\left(z,x(t)\right)}{\partial z^2} \right)^2  (E_z I_z) \, \frac{d l (z,x(t))}{dz} \, dz ,
\end{eqnarray*}
with $E_z$ and $\rho$ the Young modulus and density of the beam respectively. 
$w$ and $e$ are its width and thickness respectively. $I_z=1/12 \, w e^3$ is the corresponding second moment of area.

For our device, the beam corresponds to one foot of the structure. We have to take into account the kinetic energy due to the mass load $m_l$ (the 'paddle'), located at the end of the beam:
\begin{displaymath}
 E_{c,\,load}(t) = \frac{1}{2} m_{l} \dot{x}(t)^2.
\end{displaymath}
The model can be easily adapted to other mass load configurations.

A constant damping term per unit length $d\Lambda/dz$ (materials linear regime) enables to write the power losses due to non-conservative forces:
\begin{displaymath}
 P_n = - 2 \int_{0}^{h_z} \frac{d \Lambda}{d z} \left( \frac{\partial \, f\left(z,x(t)\right)}{\partial t} \right)^2 \frac{d l (z,x(t))}{dz}   \,  dz.
\end{displaymath}
In the most general formulation, a reactive term should also be considered. Reactive and dissipative components are linked through Kramers-Kronig relations (valid in the materials' linear regime), since they are related to the real and imaginary parts of the acoustic susceptibility of the vibrating structure  \cite{castroNeto}. With the same notations as above, we write:
\begin{displaymath}
P_n'  =  
- 2 \int_{0}^{h_z} \frac{d \Lambda'}{d z} \, \frac{\partial \, f\left(z,x(t)\right)}{\partial t}   \frac{\partial^2 \, f\left(z,x(t)\right)}{\partial t^2}  \frac{d l (z,x(t)) }{dz}  \, dz . 
\end{displaymath}
Since the end mass load is perfectly rigid, no dissipation is associated to it. The model can be adapted easily to take into account a friction occurring at the paddle.

Finally, the power of the driving force $F(t)$, applied at the point of maximal deflection $x(t)$, is:
\begin{displaymath}
P_F = \dot{x}(t)\, F(t) .
\end{displaymath}
The modeling can be adapted, of course, if the force is applied to another point of the structure, or is distributed. Note that for our devices, the driving force itself is linear, as opposed to electrostatic actuation \cite{nonlin_electrostat, superharm, geom2}.\\

The energy balance, or theorem of mechanical power, states (the structure is symmetric, we thus take twice one foot's terms):
\begin{displaymath}
\frac{d \left[ 2E_c(t)+2E_p(t)+ E_{c,\,load}(t) \right]}{dt} =  2 P_n + 2 P_n' + P_F,
\end{displaymath}
which enables to write 
 the dynamics equation for the mechanical mode under study, up to order 3 (after simplification by $\dot{x}(t)$): 
\begin{eqnarray}
 \left[ m_f\left(1+ m_1 \, x + m_2 \, x^2 \right)+ m_{l} \right] \ddot{x}  &+ & \nonumber \\
 2 \Lambda' \, \ddot{x} \left( 1+ l'_1 \, x + l'_2 \, x^2\right) + \left( \Gamma_0+ \Gamma_1 \, x \right)\, \dot{x}^2 & +& \nonumber \\ 
 2 \Lambda \, \dot{x} \left( 1+ l_1 \, x + l_2 \, x^2\right) & +& \nonumber  \\ 
 k \left( 1+ k_1 \, x + k_2 \, x^2\right) \!\!\! &  x &  \nonumber \\ 
 & = & \!\!  F(t),  \label{nonlin}
\end{eqnarray}
with $m_f$ the normal mass associated to the two feet.
This polynomial writing, within our assumptions, is the {\it most generic form} for the dynamics equation. The development is {\it unique}, and the definition of all the coefficients as a function of the mode shape $f$ is given in the Appendix.
$m_f$ and $m_l$ can be grouped together in $m = m_f+m_l$. Since the $\Lambda'$ and $m$ terms play the same role in the above equation, they can also  be grouped together. Eq. (\ref{nonlin}) can thus be rewritten, without loss of generality:
\begin{eqnarray}
  \tilde{m} \left(1+ \tilde{m}_1 \, x + \tilde{m}_2 \, x^2 \right) \ddot{x}  &+ & \nonumber \\
  + \tilde{m} \, \left( \frac{\Gamma_0}{\tilde{m}}+ \frac{\Gamma_1}{\tilde{m}} \, x \right)\, \dot{x}^2 & +& \nonumber \\ 
 2 \Lambda \, \dot{x} \left( 1+ l_1 \, x + l_2 \, x^2\right) & +& \nonumber  \\ 
 k \left( 1+ k_1 \, x + k_2 \, x^2\right) \!\!\! &  x &  \nonumber \\ 
 & = & \!\!  F(t),  \label{nonlin2}
\end{eqnarray}
with $\tilde{m}= m + 2 \Lambda'$, $\tilde{m}_1 = ( m_f m_1 + 2 \Lambda' \, l'_1 )/\tilde{m}$, and $\tilde{m}_2 = ( m_f m_2 + 2 \Lambda' \, l'_2 )/\tilde{m}$. In the following, we will drop the tilde in order to keep the writing lighter. Remember that in the Appendix the calculated $m_i$ refer to one bare foot (a cantilever or a doubly-clamped beam), without load.\\
\vspace{-1 mm}

Eq. (\ref{nonlin2}) reveals a non-linear spring force, a non-linear inertia, and a peculiar non-linear damping/inertia term with a  $\dot{x}(t)^2$ dependence.
Note that a geometrical non-linearity has intrinsically a {\it similar impact on both} the inertia and the restoring force of a device. The damping appears also to be intrinsically non-linear.
$m$, $k$ and $\Lambda$ are the normal mass, normal spring constant and dissipation constant relative to the mode's linear regime. 

$m_i$ (inertial), $k_i$ (elastic), $l_i$ and $\Gamma_i$ (damping) are the non-linear constants arising from the exact shape of the (dynamical) distortion $f$ of the structure under study.
There are constraints on these coefficients, and some can be proven to be irrelevant to our experiments, leading to the simplified Eq. (\ref{dyneq}), Section \ref{exp} (see Appendix). 
However for the sake of completeness, we give below the full mathematical solutions to Eq. (\ref{nonlin2}) in the two experimental conditions of interest to us: harmonic-drive (with an extension of Landau \& Lifshitz's method) and free-decay (with an application of Lindstedt \& Poincar\'e's method).
Note that any additional non-linear effect preserving the analytic shape of Eq. (\ref{nonlin2}), like an air drag force $\propto \dot{x}^2$, can be taken into account by our fitting solutions.

\subsection{Landau-Lifshitz method}

This theoretical technique gives the exact solution of Eq. (\ref{nonlin2}) in the case $F(t)=F_0 \cos(\omega t)$.
The original method of Landau \& Lifshitz \cite{landaumeca} considers only $k_1, k_2 \neq 0$, with a steady state attained for $F_0=0$ and no damping ($\Lambda=0$). See also \cite{book2} for a good discussion of the method.
We extend here the theory from \cite{JLTP_VIW} using notations of the present article.
Following Landau \& Lifshitz, we postulate for the solution:
\begin{displaymath}
x(t) = \sum_{n=0}^{+\infty} a^c_{n}(\omega) \cos (n\,\omega t) + \sum_{n=1}^{+\infty} a^s_{n}(\omega) \sin (n\,\omega t)
\end{displaymath}
and seek only the static term $n=0$, plus the first harmonic $n=1$. In \cite{JLTP_VIW}, higher orders where simply taken to be zero; here, we also retain $a^c_{2}, a^s_{2}$. We define $x_0=\sqrt{(a^c_{1})^2+(a^s_{1})^2}$ the amplitude of the first harmonic displacement.

The solution is a simple {\it modified Lorentzian}:
\begin{eqnarray*}
a^c_{0} & = & \beta_0 \, x_0^2 , \\
a^c_{1}(\omega) & = & \frac{F_{0}}{k} \frac{(\omega_r/\omega_{0})^2-(\omega/\omega_{0})^2}{\left[(\omega_r/\omega_{0})^2-(\omega/\omega_0)^2 \right]^2+ \left[\Delta \omega \, \omega/(\omega_0)^2\right]^2}, \\
a^s_{1}(\omega) & = & \frac{F_{0}}{k} \frac{\Delta \omega \, \omega/(\omega_0)^2}{\left[(\omega_r/\omega_{0})^2-(\omega/\omega_0)^2 \right]^2+ \left[\Delta \omega \, \omega/(\omega_0)^2\right]^2}.
\end{eqnarray*}
In these expressions, the resonance position $\omega_r$ and the resonance linewidth term $\Delta \omega$ are now functions of $x_0$:
\begin{eqnarray*}
\omega_r & = & \sqrt{\omega_0^2 + 2 \beta_1 \omega_0 \, x_0^2} \, \approx \, \omega_0 + \beta_1 \, x_0^2 ,\\
\Delta \omega & = & \Delta \omega_0 + \beta_2 \, x_0^2 ,
\end{eqnarray*}
with the usual definitions $\omega_0=\sqrt{k/m}$, and $\Delta \omega_0 =2 \Lambda/m$ (expressed in Rad/s). 
The maximal displacement amplitude is obtained for:
\begin{equation}
\omega_{res}^2 = \omega_0^2 \left[ 1 - 1/(2 Q^2) \right] + 2 \omega_0  \left[ \beta_1 - \beta_2/ (2 Q) \right] x_0^2  \label{maxif}
\end{equation}
with $Q=\omega_0/\Delta \omega $. The calculation brings:
\begin{eqnarray}
\beta_0 & = & -\frac{1}{2} \left[ k_1 + \left( \frac{\omega}{\omega_0} \right)^2  \left( \frac{\Gamma_0}{m} - m_1 \right) \right], \nonumber \\
\beta_1 & = & 
  + \frac{\omega_0}{2} \left[ \frac{3}{4} k_2 - \left( \frac{\omega}{\omega_0} \right)^2 \left( \frac{3}{4} m_2 -\frac{1}{4}  \frac{\Gamma_1}{m} \right) + \right. \nonumber \\
&& \!\!\!\!\!\!\!\!\!\!\!\!\!\!\!\!\!\!\!\! \left.  \left( \frac{1}{2} m_1 \left( \frac{\omega}{\omega_0} \right)^2 -k_1 \right) \left( k_1 + \left( \frac{\omega}{\omega_0} \right)^2 \left( \frac{\Gamma_0}{m} - m_1 \right) \right) \right. \nonumber \\
&& \!\!\!\!\!\!\!\!\!\!\!\!\!\!\!\!\!\!\! \left. + \frac{(4 \, \omega^2 - \omega_0^2)\,\omega_0^2}{(4\, \omega \Lambda/ m )^2+(4 \, \omega^2 - \omega_0^2)^2} \times \right. \nonumber \\
&& \!\!\!\!\!\!\!\!\!\!\!\!\!\!\!\!\!\!\!\!\!\!\!\!\!\!\!\!\!\!\!\!\!\!\!\!\!\!\!\! \left. \left(\frac{1}{2} \left( k_1  
  - \frac{1}{2}\left( \frac{\omega}{\omega_0} \right)^2 \left( 5\, m_1 - 4 \frac{\Gamma_0}{m} \right) \right) \!\! \left( k_1 - \left( \frac{\omega}{\omega_0} \right)^2 \left( \frac{\Gamma_0}{m} + m_1 \right) \right) \right. \right. \nonumber \\
 && \!\!\!\!\!\!\!\!\!\!\!\!\!\!\!\!\!\!\!\! \left. \left.-  \left( \frac{\omega \Lambda/m}{\omega_0^2}\right)^{2} \, l_1^2 \right) + 
 \frac{(4 \, \omega \Lambda / m)^2}{(4\, \omega \Lambda/ m )^2+(4 \, \omega^2 - \omega_0^2)^2} \, \frac{3\, l_1}{8} \times  \right. \nonumber \\
 && \left. \left( - k_1  
  +\left( \frac{\omega}{\omega_0} \right)^2 \left( 2\, m_1 -  \frac{\Gamma_0}{m} \right) \right) \right],  \label{beta1} \\
\beta_2 & = & +\frac{\Lambda}{m}\left[ \frac{1}{2}l_2 - l_1 \left( k_1+ \left( \frac{\omega}{\omega_0} \right)^2 \left(  \frac{\Gamma_0}{m} - m_1 \right)\right) \right. \nonumber \\
&& \!\!\!\!\!\!\!\!\!\!\!\!\!\!\!\!\!\!\!\! \left. + \frac{(4 \, \omega^2 - \omega_0^2)\,\omega_0^2}{(4\, \omega \Lambda/ m )^2+(4 \, \omega^2 - \omega_0^2)^2} \, l_1 \times \right. \nonumber \\
&& \!\!\!\!\!\!\!\!\!\!\!\!\!\!\!\!\!\!\!\!\!\!\!\!\!\!\!\!\!\!\!\!\!\!\!\!\!\!\!\! \left. \left( \frac{1}{2}\left( k_1 - \left( \frac{\omega}{\omega_0} \right)^2 \left( \frac{\Gamma_0}{m} + m_1 \right) \right)+ k_1  
  - \frac{1}{2}\left( \frac{\omega}{\omega_0} \right)^2 \left( 5\, m_1 - 4 \frac{\Gamma_0}{m} \right) \right) \right. \nonumber \\
&& \!\!\!\!\!\!\!\!\!\!\!\!\!\!\!\!\!\!\! + \left. \frac{4\,\omega_0^4}{(4\, \omega \Lambda/ m )^2+(4 \, \omega^2 - \omega_0^2)^2} \times \right. \nonumber \\
&& \!\!\!\!\!\!\!\!\!\!\!\!\!\!\!\!\!\!\!\!\!\!\!\!\!\!\!\!\!\!\!\!\!\!\!\!\!\!\!\! \left. \left(\frac{1}{2} \left( k_1  
  - \frac{1}{2}\left( \frac{\omega}{\omega_0} \right)^2 \left( 5\, m_1 - 4 \frac{\Gamma_0}{m} \right) \right) \!\! \left( k_1 - \left( \frac{\omega}{\omega_0} \right)^2 \left( \frac{\Gamma_0}{m} + m_1 \right) \right) \right. \right. \nonumber \\
  && \left. \left.-  \left( \frac{\omega \Lambda/m}{\omega_0^2}\right)^{2} \,  l_1^2 \right) \right] \nonumber.
\end{eqnarray}
The above is valid for {\it any $Q$}.
Finding $a^c_{1},a^s_{1}$ from the above expressions reduces to find the roots of a simple polynom, with $y=x_0^2$:
\begin{eqnarray*}
& - & \left( \frac{F_0}{k}\right) ^2  \omega_0^4 + \left(\Delta \omega_0^2 \omega^2+(\omega^2-\omega_0^2)^2 \right)\, y \\
& + & \left(  2 \beta_2 \, \Delta \omega_0 \omega^2 - 4 \beta_1 \, \omega_0 (\omega^2-\omega_0^2) \right)\, y^2 \\
& + & \left( \beta_2^2 \, \omega^2 + 4 \beta_1^2 \, \omega_0^2\right) \, y^3=0.
\end{eqnarray*}
There are three roots for $y$, which can be found {\it analytically} (see Supplemental Material). One then replaces $x_0^2$ in the expressions of $a^c_{1},a^s_{1}$.
Below  a critical oscillation amplitude $x_c$, only one root is physical (real positive). Above $x_c$, three branches coexist: two physical solutions (plus a metastable branch), corresponding to upwards and downwards frequency sweeps \cite{landaumeca}.

In Section \ref{exp}, we work in the high $Q$ limit (meaning in particular $\omega \approx \omega_0$) with also $\beta_2=0$, and define $\beta=\beta_1/\omega_0$. The expressions above are given for $x_0$ defined as a {\it peak amplitude}. 
The frequency is pulled quadratically with the amplitude $x_0$, Fig. \ref{quadfreq}, Eq. (\ref{maxif}), with $\omega_{res} \approx \omega_r$.
With $\beta_2=0$, when sweeping the resonance in the direction of the non-linear coefficient $\beta_1$ (i.e. upwards for positive), the height of the detected peak is still proportional to the applied force, and inversely proportional to the damping term, through the simple relation $x_0=F_0/k \, Q$, Fig. \ref{elast} \cite{coating, JLTP_VIW}. We have $x_c = \frac{2}{3} 3^{1/4} \sqrt{\omega_0/(Q\, \left|\beta_1\right|)}$ \cite{landaumeca}.

\subsection{Lindstedt-Poincar\'e method}

Take Equation (\ref{nonlin2}) with $F(t)=0$ and divide it by the non-linear 'mass' $m\left(1+ m_1 \, x + m_2 \, x^2 \right) $. The new equation developed to order 3 in $x$ contains the non-linear parameters:
\begin{eqnarray*}
\lambda_a & = & \left( k_1 - m_1 \right) , \\
\lambda_b & = & \left(k_2 - k_1 m_1 + m_1^2 - m_2 \right), \\
\lambda_c & = & \left( l_1 - m_1 \right) , \\
\lambda_d & = & \left( l_2 - l_1 m_1 + m_1^2 - m_2  \right) , \\
\lambda_e & = & \frac{\Gamma_0}{m}, \\
\lambda_f & = & \left( \frac{\Gamma_1}{m} - \frac{\Gamma_0}{m} m_1 \right). 
\end{eqnarray*}
Note that $\lambda_a, \lambda_c, \lambda_e$ are homogeneous to m$^{-1}$, while $\lambda_b, \lambda_d, \lambda_f$ to m$^{-2}$.
We apply the Lindstedt-Poincar\'e method of solving non-linear differential equations \cite{lindstedt,book2}. Let us write the sought solution in the form of a series: 
\begin{eqnarray*}
x(t) & = & s_0(t) \, + \\
&&   \lambda_a s_{a,1}(t) + \lambda_b s_{b,1}(t) +\lambda_c s_{c,1}(t) + \\
&&   \lambda_d s_{d,1}(t)+\lambda_e s_{e,1}(t)+\lambda_f s_{f,1}(t) + \\
&&   \lambda_a^2 s_{a,2}(t) +\lambda_c^2 s_{c,2}(t) +\lambda_e^2 s_{e,2}(t) +\\
&&   \lambda_{a}\lambda_{c} s_{a,c}(t)+ \lambda_{a}\lambda_{e} s_{a,e}(t)+ \lambda_{c} \lambda_{e} s_{c,e}(t) +..., 
\end{eqnarray*}
expanded here at second order.
The Lindstedt-Poincar\'e method is based on the idea that the function $s_0(t)$ should involve an oscillation frequency $\omega_R$ which is also a series of the $\lambda_i$, written in such a way that the expansion of the function {\it should cancel all secular terms} which would remain in a standard perturbative theory \cite{lindstedt} (see below). We thus introduce:
\begin{eqnarray*}
\omega_R &=& \omega_0 + \\
&&\lambda_a \, \omega_{a} + \lambda_b \, \omega_{b}+ \lambda_c \, \omega_{c} + \lambda_d \, \omega_{d}
+ \lambda_e \, \omega_{e} + \lambda_f \, \omega_{f}+ \\
&& \lambda_a^2 \, \omega_{a,2}+ \lambda_c^2 \, \omega_{c,2}+ \lambda_e^2 \, \omega_{e,2}+ \\
&&  \lambda_{a}\lambda_{c} \, \omega_{a,c}+ \lambda_{a}\lambda_{e} \, \omega_{a,e}+ \lambda_{c} \lambda_{e} \, \omega_{c,e}+...
\end{eqnarray*}
written here at the lowest compatible order, with $\omega_{a}$ to $\omega_{c,e}$ the unknowns to be defined. By construction, the solution $s_0$ is:
\begin{displaymath}
s_0(t) = X_0 \exp \left(-  \frac{\omega_R}{\omega_0} \frac{t}{\tau} \right) \cos \left(\sqrt{1-\left(\frac{1}{\omega_0 \tau}\right)^2} \omega_R \, t + \varphi \right)
\end{displaymath}
with $\tau=2/\Delta \omega_0$ the ring-down time ($\omega_0=\sqrt{k/m}$, $\Delta \omega_0 =2 \Lambda/m$, $X_0$ and $\varphi$ are initial conditions). 
In the canonical Lindstedt-Poincar\'e problem, the damping is zero ($\Lambda=0$) and only $k_2$ (thus here $\lambda_b$) is taken into account. 
In this case, the perturbative theory brings:
\begin{displaymath}
\ddot{s_b}(t)+ \omega_0^2 \,  s_b(t)= -\frac{3}{4} X_0^3 \omega_0^2 \, \cos \left( \omega_0 t + \varphi \right),
\end{displaymath}
which particular solution is $s_b(t)=-3/8 \, X_0^3 \,(\omega_0  t) \,\times$ $ \sin \left( \omega_0 t + \varphi \right)$, a function with divergent amplitude at large $t$ called the {\it secular term}, which can be canceled by the appropriate choice $\omega_b=+3/8 \, X_0^2 \,\omega_0$. 

With a non-zero damping (and all $\lambda_i$ taken into account), the 'pathologic' equations appearing in the resolution rewrite, with the change of argument $t \rightarrow \omega_R t$:
\begin{eqnarray*}
&& \frac{d^2 \, s_i(\omega_R t)}{d (\omega_R t)^2}+ \frac{ \Delta \omega_0 }{\omega_0} \, \frac{d \, s_i(\omega_R t)}{d (\omega_R t)}+  s_i(\omega_R t)  =  \\
&& A_i \, \exp \left( - 3 \frac{\omega_R t}{\omega_0 \tau} \right) \, \cos \left( \sqrt{1-\left(\omega_0 \tau\right)^{-2}} \omega_R t + \phi_i \right),
\end{eqnarray*}
with $A_i$ and $\phi_i$ defined through $X_0$ and $\varphi$ (and $i = b;d;f;a,2;c,2;e,2;a,c;a,e;c,e$).
The particular solution is again analytic:
\begin{eqnarray*}
&& s_i(\omega_R t)  =   \\
&& + \frac{A_i}{4} \exp \left( - 3 \frac{\omega_R t}{\omega_0 \tau} \right) \, \left[ \cos \left( \sqrt{1-\left(\omega_0 \tau \right)^{-2}} \omega_R t + \phi_i \right)  \right. \\
&& \left. - \left( \omega_0 \tau \right) \sqrt{1-\left(\omega_0 \tau \right)^{-2}} \sin \left( \sqrt{1-\left(\omega_0 \tau \right)^{-2}} \omega_R t + \phi_i \right) \right]  \\
&&+ \frac{A_i}{4} \exp \left( - \frac{\omega_R t}{\omega_0 \tau} \right) \, \left( \omega_0 \tau \right) \sin \left( \sqrt{1-\left(\omega_0 \tau \right)^{-2}} \omega_R t + \phi_i \right)
\end{eqnarray*}
but this time, it is {\it not} pathologic. 
However in the high $Q$ limit ($Q=\omega_0 \tau/2$, thus ($\omega_0 \tau)^{-1}  \rightarrow 0$), the above expression produces the secular solution. 
Note that the problem is perfectly regular in the vicinity of $\Lambda \rightarrow 0$, so the idea behind our calculation is that 
Lindstedt \& Poincar\'e's approach is still valid: these functions should be canceled by the proper choice of $\omega_i$ in the final solution.

The first order terms are not pathologic and simply bring $\omega_a=\omega_c=\omega_e=0$. 
As opposed to the standard method, the higher order $\omega_i$ will turn out to be time-dependent functions. The equations they are involved in write:
\begin{eqnarray*}
&& \frac{\omega_b(t)+ t \, \dot{\omega_b}(t)}{\omega_0}   +  \\
&& \left[ \frac{\tan \left(\sqrt{1-\left(\omega_0 \tau \right)^{-2}} \omega_R \, t + \varphi \right)}{2} \right] \frac{2 \dot{\omega_b} (t) + t \, \ddot{\omega_b}(t)}{\omega_0^2}  =  \\
&& \!\!\!\!\!\!\!\!\!\!\!\! + \frac{3}{8} \left( X_0 \right)^2 \exp \left(- 2 \frac{t}{\tau} \right) \left[1 + \frac{\tan \left(\sqrt{1-\left(\omega_0 \tau \right)^{-2}} \omega_R \, t + \varphi \right)}{\omega_0 \tau} \right], \\
&& \frac{\omega_d(t)+ t \, \dot{\omega_d}(t)}{\omega_0}   +  \\
&& \left[ \frac{\tan \left(\sqrt{1-\left(\omega_0 \tau \right)^{-2}} \omega_R \, t + \varphi \right)}{2} \right] \frac{2 \dot{\omega_d} (t) + t \, \ddot{\omega_d}(t)}{\omega_0^2}  =  \\
&& \!\!\!\!\!\! -\frac{1}{4} \left( X_0 \right)^2 \exp \left(- 2 \frac{t}{\tau} \right) \left[ \frac{\tan \left(\sqrt{1-\left(\omega_0 \tau \right)^{-2}} \omega_R \, t + \varphi \right)}{\omega_0 \tau} \right],\\
&& \frac{\omega_f(t)+ t \, \dot{\omega_f}(t)}{\omega_0}   +  \\
&& \left[ \frac{\tan \left(\sqrt{1-\left(\omega_0 \tau \right)^{-2}} \omega_R \, t + \varphi \right)}{2} \right] \frac{2 \dot{\omega_f} (t) + t \, \ddot{\omega_f}(t)}{\omega_0^2}  =  \\
&& \!\!\!\!\!\!\!\!\!\!\!\!\!\! + \frac{1}{8} \left( X_0 \right)^2 \exp \left(- 2 \frac{t}{\tau} \right) \left[1 + \frac{3 \tan \left(\sqrt{1-\left(\omega_0 \tau \right)^{-2}} \omega_R \, t + \varphi \right)}{\omega_0 \tau} \right] \!\! ,
\end{eqnarray*}
and:
\begin{eqnarray*}
&& \frac{\omega_{a,2}(t)+ t \, \dot{\omega_{a,2}}(t)}{\omega_0}   +  \\
&& \left[ \frac{\tan \left(\sqrt{1-\left(\omega_0 \tau \right)^{-2}} \omega_R \, t + \varphi \right)}{2} \right] \frac{2 \dot{\omega_{a,2}} (t) + t \, \ddot{\omega_{a,2}}(t)}{\omega_0^2}  =  \\
&&\!\!\!\!\!\!\!\!\!\!\!\! - \frac{5}{12} \left( X_0 \right)^2 \exp \left(- 2 \frac{t}{\tau} \right) \left[1 + \frac{11 \tan \left(\sqrt{1-\left(\omega_0 \tau \right)^{-2}} \omega_R \, t + \varphi \right)}{15 \, \omega_0 \tau} \right], \\
&& \frac{\omega_{c,2}(t)+ t \, \dot{\omega_{c,2}}(t)}{\omega_0}   +  \\ 
&& \left[ \frac{\tan \left(\sqrt{1-\left(\omega_0 \tau \right)^{-2}} \omega_R \, t + \varphi \right)}{2} \right] \frac{2 \dot{\omega_{c,2}} (t) + t \, \ddot{\omega_{c,2}}(t)}{\omega_0^2}  = 
\end{eqnarray*}
\begin{eqnarray*}
&& \!\!\!\!\!\!\!\!\!\!\!\!\!\!\!\!\!\! - \frac{1}{6} \frac{\left( X_0 \right)^2}{\left( \omega_0 \tau \right)^2} \exp \left(- 2 \frac{t}{\tau} \right) \left[1 + \frac{13 \tan \left(\sqrt{1-\left(\omega_0 \tau \right)^{-2}} \omega_R \, t + \varphi \right)}{3\, \omega_0 \tau} \right], \\ 
&& \frac{\omega_{e,2}(t)+ t \, \dot{\omega_{e,2}}(t)}{\omega_0}   +  \\
&& \left[ \frac{\tan \left(\sqrt{1-\left(\omega_0 \tau \right)^{-2}} \omega_R \, t + \varphi \right)}{2} \right] \frac{2 \dot{\omega_{e,2}} (t) + t \, \ddot{\omega_{e,2}}(t)}{\omega_0^2} =  \\
&& \!\!\!\!\!\!\!\!\!\!\!\!\!\!\!\!\!\! - \frac{1}{6} \left( X_0 \right)^2 \exp \left(- 2 \frac{t}{\tau} \right) \left[1 + \frac{19 \tan \left(\sqrt{1-\left(\omega_0 \tau \right)^{-2}} \omega_R \, t + \varphi \right)}{3\, \omega_0 \tau} \right],   
\end{eqnarray*}
and finally:
\begin{eqnarray*}
&& \frac{\omega_{a,c}(t)+ t \, \dot{\omega_{a,c}}(t)}{\omega_0}   +  \\
&& \left[ \frac{\tan \left(\sqrt{1-\left(\omega_0 \tau \right)^{-2}} \omega_R \, t + \varphi \right)}{2} \right] \frac{2 \dot{\omega_{a,c}} (t) + t \, \ddot{\omega_{a,c}}(t)}{\omega_0^2}  =  \\
&& \!\!\!\!\!\! +\frac{1}{4} \left( X_0 \right)^2 \exp \left(- 2 \frac{t}{\tau} \right) \left[ \frac{\tan \left(\sqrt{1-\left(\omega_0 \tau \right)^{-2}} \omega_R \, t + \varphi \right)}{\omega_0 \tau} \right], \\
&& \frac{\omega_{a,e}(t)+ t \, \dot{\omega_{a,e}}(t)}{\omega_0}   +  \\ 
&& \left[ \frac{\tan \left(\sqrt{1-\left(\omega_0 \tau \right)^{-2}} \omega_R \, t + \varphi \right)}{2} \right] \frac{2 \dot{\omega_{a,e}} (t) + t \, \ddot{\omega_{a,e}}(t)}{\omega_0^2}  = \\
&& \!\!\!\!\!\!\!\!\!\!\!\!\!\!\!\!\!\!\! - \frac{5}{12} \left( X_0 \right)^2 \exp \left(- 2 \frac{t}{\tau} \right) \left[1 + \frac{41 \tan \left(\sqrt{1-\left(\omega_0 \tau \right)^{-2}} \omega_R \, t + \varphi \right)}{15 \, \omega_0 \tau} \right] \!, \\
&& \frac{\omega_{c,e}(t)+ t \, \dot{\omega_{c,e}}(t)}{\omega_0}   +  \\
&& \left[ \frac{\tan \left(\sqrt{1-\left(\omega_0 \tau \right)^{-2}} \omega_R \, t + \varphi \right)}{2} \right] \frac{2 \dot{\omega_{c,e}} (t) + t \, \ddot{\omega_{c,e}}(t)}{\omega_0^2} = \\
&&  \!\!\!\!\!\! +\frac{1}{4} \left( X_0 \right)^2 \exp \left(- 2 \frac{t}{\tau} \right) \left[ \frac{\tan \left(\sqrt{1-\left(\omega_0 \tau \right)^{-2}} \omega_R \, t + \varphi \right)}{\omega_0 \tau} \right] \!.
\end{eqnarray*}
The terms in brackets involving the tangent function have been developed at first order in $1/(\omega_0 \tau)$.
The second term $(2 \dot{\omega_i} (t) + t \, \ddot{\omega_i}(t))/(\omega_0^2)$ in the left hand side of these equations is $1/(\omega_0 \tau)$ smaller than the first one. Solving at first order in $1/(\omega_0 \tau)$ (the high $Q$ limit) is straightforward:
\begin{eqnarray*}
\omega_b(t) & = & + \frac{3}{8} \omega_0 \, \left( X_0 \right)^2 \, E(t), \\
\omega_d(t) & = & 0, \\
\omega_f(t) & = & + \frac{1}{8} \omega_0 \, \left( X_0 \right)^2 \, E(t), \\
\omega_{a,2}(t) & = & - \frac{5}{12} \omega_0 \, \left( X_0 \right)^2 \, E(t),\\
\omega_{c,2}(t) & = & 0, \\
\omega_{e,2}(t) & = & - \frac{1}{6} \omega_0 \, \left( X_0 \right)^2 \, E(t), \\
\omega_{a,c}(t) & = & 0, \\
\omega_{a,e}(t) & = & - \frac{5}{12} \omega_0 \, \left( X_0 \right)^2 \, E(t),\\
\omega_{c,e}(t) & = & 0, 
\end{eqnarray*}
with $E(t)=\left[ 1 - \exp \left(- t/(\tau/2) \right)\right]/\left[t/(\tau/2) \right] $. Regrouping all terms, we obtain:
\begin{displaymath}
\omega_R  =  \omega_0 + \lambda_1 \,X_0^2\, E(t) . 
\end{displaymath}
The total perturbative solution brings oscillating terms at $2 \omega_R$ and  $3 \omega_R$, plus a 'constant', all decaying with exponential prefactors. Writing only the latter, we have:
\begin{displaymath}
x(t)  =  s_0(t) \, + \lambda_0 \, X_0^2 \, \exp \left(- 2 \frac{t}{\tau} \right).
\end{displaymath}
The global non-linear coefficients are:
\begin{eqnarray}
& & \lambda_0  =  -\frac{1}{2} \left[ \left( k_1 - m_1 \right) + \frac{\Gamma_0}{m} \right], \nonumber \\
& & \!\!\!\!\!\!\!\! \lambda_1  =  \omega_0 \,\! \left[ \frac{3}{8} \left(k_2 - k_1 m_1 + m_1^2 - m_2 \right) + \frac{1}{8} \left( \frac{\Gamma_1}{m} - \frac{\Gamma_0}{m} m_1 \right)   \right. \nonumber \\
& - & \left. \frac{5}{12} \left( k_1-m_1 \right)^2  - \frac{1}{6} \left( \frac{\Gamma_0}{m} \right)^2  -\frac{5}{12} \left( k_1-m_1 \right)\frac{\Gamma_0}{m}  \right] \label{lambda1} \! .
\end{eqnarray}
In Section \ref{exp}, we define $\lambda=\lambda_1/\omega_0$. The expressions above are given for $X_0$ defined as a {\it peak amplitude}. 

Note that the whole calculation has been done assuming the high $Q$ limit. 
Comparing the two resolutions pushed at equivalent order (harmonic drive and free decay), we realize that $\beta_0 = \lambda_0$, and $\beta_1 = \lambda_1$ in this limit. 
While the measured damping is also non-linear with $\beta_2$ in the harmonic solution, the free-decay function (in the high $Q$ limit) presents a constant relaxation time $\tau$.

\section{CONCLUSIONS}

In conclusion, we presented two types of measurements of the geometrical non-linear behavior of cantilever MEMS, a frequency-sweep and a free-decay technique. Two theoretical expressions are presented, based on the Landau-Lifshitz and Lindstedt-Poincar\'e methods.
Fits enable to extract the oscillators' characteristics, in the whole dynamic range from a linear to a very non-linear regime. 
We show that using these expressions non-linear coefficients can be obtained experimentally.
The novelty of the approach lies in its {\it generic and purely analytic nature}.
The experimental and theoretical methods presented enable a characterization of MEMS, NEMS, or any other weakly geometrically non-linear mechanical oscillator described by a single degree of freedom beyond the simple Duffing model.
The detailed understanding of the geometrical non-linear behavior is also a basis for further studies with {\it anelastic} materials, and more refined properties of non-linear dynamics.

\begin{acknowledgments}
We wish to thank T. Fournier, C. Lemonias, and B. Fernandez for their help in the fabrication of samples, and  
J. Parpia for valuable discussions.
We acknowledge the support from MICROKELVIN, the EU FRP7 low temperature infrastructure grant 228464. \\
\vspace*{10mm}

Note: {\it Mathematica} codes can be produced as Supplemental Material upon request.
\vspace*{10mm}

\end{acknowledgments}

\begin{appendix}
\section{Defining the non-linear coefficients from the mode shape}

In Section \ref{theory} we give the generic non-linear dynamics equation of a geometrically non-linear 1D oscillator, Eq. (\ref{nonlin}). The expansion is based on the mode shape $f$, which is a well-defined and regular given function. In the present Appendix, we give all the non-linear coefficients that have been introduced for a cantilever structure or a doubly-clamped beam. If the function $f$ is obtained from a specific theory, one can then explicitly calculate the non-linear coefficients.\\

The linear parameters introduced in Section \ref{theory} are defined as:
\begin{eqnarray*}
m & = & m_f + m_l , \\
m_f & = & \rho \, w e\,( 2  m_0), \\
k & = & E_z I_z \, ( 2 k_0), \\
\Lambda & = & \frac{d \Lambda}{d z} \, (2 m_0),\\
\Lambda' & = & \frac{d \Lambda'}{d z} \,(2 m_0).
\end{eqnarray*}

The non-linear parameters introduced for the beam length are:
\begin{eqnarray*}
h_2 & = &  +\frac{1}{2} \int_0^h \left( \frac{\partial^2 f(z,x=0) }{\partial z \partial x } \right)^2 dz/h, \\
h_3 & = & +\frac{1}{2} \int_0^h    \frac{\partial^2 f(z,x=0) }{\partial z \partial x } \frac{\partial^3 f(z,x=0) }{\partial z \partial x^2 } \, dz/h.
\end{eqnarray*}

The quadratic-velocity terms are linked to the others through:
\begin{eqnarray*}
\Gamma_0 & = & \left( m_f +  2 \frac{d \Lambda'}{d z} \, ( 2 m_0 ) \right) \, \frac{1}{2} \, m_1, \\
\Gamma_1 & = & m_f\, m_2 +  2 \frac{d \Lambda'}{d z} \, (2 \gamma) .
\end{eqnarray*}

In the above definitions, the factor 2 in front of $m_0$, $k_0$ and $\gamma$ is due to the fact that our structure has two identical cantilever feet.
For a single cantilever with an end load, or a doubly-clamped beam with a mass load positioned in the middle, just remove this 2.

The calculation brings that the damping terms $l_i$ and $l'_i$ are simply equal to the $m_i$.
We are thus left with the definition of the $m_i$, $k_i$ and $\gamma$.

After performing the integrations and the series expansions, we obtain:
\begin{eqnarray*}
m_0 & = & \int_0^h \left( \frac{\partial f(z,x=0) }{\partial x }\right)^2 dz , \\
m_1 & = & \left[ \int_0^h 2\frac{\partial f(z,x=0) }{\partial x }\frac{\partial^2  f(z,x=0) }{\partial x^2 } dz \right]/ \, m_0, \\ 
\end{eqnarray*}
and:
\begin{eqnarray*}
k_0 & = & \int_0^h \left(\frac{\partial^3 f(z,x=0) }{\partial z^2 \partial x }\right)^2 dz, \\
k_1 & = & \frac{3}{2} \left[ \int_0^h \frac{\partial^3{f(z,x=0)}}{\partial z^2 \partial x }\frac{\partial^4 f(z,x=0) }{\partial z^2 \partial x^2 }dz \right]/ \, k_0, \\
\end{eqnarray*}
and finally:
\begin{displaymath}
\!\!\!\!
\gamma   =   \int_0^h \left[ \left( \frac{\partial^2 f(z,x=0) }{\partial x^2 } \right)^2 + \frac{\partial f(z,x=0) }{\partial x }\frac{\partial^3 f(z,x=0) }{\partial x^3 } \right] dz	.
\end{displaymath}
Only the two coefficients $m_2$ and $k_2$ differ for cantilever and doubly clamped beams.

\subsection{Cantilever}

The two second order coefficients are:
\begin{eqnarray*}
m_ 2 & = & \int_0^h \left[  \left(\frac{\partial^2 f(z,x=0) }{\partial x^2 }\right)^2 \right. \\  
     & + &  \left.  \frac{\partial f(z,x=0) }{\partial x }\frac{\partial^3 f(z,x=0) }{\partial x^3 } \right.   \\ 
     & + &  \left.  \frac{1}{2}  \left(\frac{\partial f(z,x=0) }{\partial x }\right)^2 
                                 \left(\frac{\partial^2 f(z,x=0) }{\partial z \partial x }\right)^2 \right] dz / \, m_0 \\ 
     & - & \!\! \frac{1}{2}  \int_0^h  \!\! \left(\frac{\partial^2 f(z,x=0) }{\partial z \partial x }\right)^2 \!\!\! dz \!  \left(\frac{\partial f(z=h,x=0) }{\partial x }\right)^2 \!\!\!  / \, m_0, \\
k_2  & = &  2 \int_0^h \left[ \frac{1}{2} \left(\frac{\partial^2 f(z,x=0) }{\partial z \partial x }\right)^2
                                          \left(\frac{\partial^3 f(z,x=0) }{\partial z^2 \partial x }\right)^2 \right. \\
& + & \left. \frac{1}{4} \left(\frac{\partial^4 f(z,x=0) }{\partial z^2 \partial x^2 }\right)^2 \right.   \\
& + & \left. \frac{1}{3} \frac{\partial^3 f(z,x=0) }{\partial z^2 \partial x } \frac{\partial^5 f(z,x=0) }{\partial z^2 \partial x^3 } \right] dz / \, k_0  \\
& - &   \int_0^h \!\! \left(\frac{\partial^2{f(z,x=0)}}{\partial{z}\partial{x}}\right)^2 \!\! dz  \! \left(\frac{\partial^3{f(z=h,x=0)}}{\partial{z^2}\partial{x}}\right)^2  \!\! / \, k_0.
\end{eqnarray*}

\subsection{Doubly-clamped beam} 

Similarly to the previous section, we obtain:
\begin{eqnarray*}
m_ 2 & = & \int_0^h \left[  \left(\frac{\partial^2 f(z,x=0) }{\partial x^2 }\right)^2 \right. \\  
     & + &  \left.  \frac{\partial f(z,x=0) }{\partial x }\frac{\partial^3 f(z,x=0) }{\partial x^3 } \right.   \\ 
     & + &  \left.  \frac{1}{2}  \left(\frac{\partial f(z,x=0) }{\partial x }\right)^2 
                                 \left(\frac{\partial^2 f(z,x=0) }{\partial z \partial x }\right)^2 \right] dz / \, m_0 , \\
k_2  & = & 2 \int_0^h \left[ \frac{1}{2} \left(\frac{\partial^2 f(z,x=0) }{\partial z \partial x }\right)^2
                                          \left(\frac{\partial^3 f(z,x=0) }{\partial z^2 \partial x }\right)^2 \right. \\
& + & \left. \frac{1}{4} \left(\frac{\partial^4 f(z,x=0) }{\partial z^2 \partial x^2 }\right)^2 \right.   \\
&+& \left. \frac{1}{3} \frac{\partial^3 f(z,x=0) }{\partial z^2 \partial x } \frac{\partial^5 f(z,x=0) }{\partial z^2 \partial x^3 } \right] dz / \, k_0 .
\end{eqnarray*}

\subsection{Reduction of equations} 

As previously stated, $l_i$ and $l'_i$ are simply equal to the $m_i$.
Moreover, no reactive contribution $\Lambda'$ could be detected experimentally for our MEMS \cite{coating}. We can thus drop the reactive contribution in the $\Gamma_i$, and in the definition of the tilded coefficients $\tilde{m}$ and $\tilde{m}_i$ (Section \ref{theory}). Thus:
\begin{eqnarray}
\tilde{m} & = & m, \nonumber \\
\Gamma_0 & = & m  \, \frac{1}{2} \, \tilde{m}_1, \nonumber\\
\Gamma_1 & = & m\, \tilde{m}_2 , \nonumber\\
\tilde{m}_i & = & \frac{m_f}{m_f+m_l} m_i \label{mis}.
\end{eqnarray}

Furthermore, the damping terms $l_i$ have a negligible impact on the resonance of our MEMS devices \cite{JLTP_VIW}, 
since in Fig. \ref{elast} no anomalous non-linear damping is visible. This can be easily understood from Section \ref{theory}, comparing the coefficients $\beta_1/\omega_0$ and $\beta_2/\Delta \omega_0$: these two terms should be roughly of the same order. In practice, the frequency shifts measured are always smaller than $1~$\% (Fig. \ref{quadfreq}), thus the linewidth non-linearity is expected to be also in the \% range. Since our linewidth fittings never resolve better than typ. $5~$\%, the $l_i$ can be safely neglected in practice. We are thus left with:
\begin{eqnarray}
  \tilde{m} \left(1+ \tilde{m}_1 \, x + \tilde{m}_2 \, x^2 \right) \ddot{x}  &+ & \nonumber \\
  + \tilde{m} \, \left( \frac{1}{2} \, \tilde{m}_1 + \tilde{m}_2 \, x \right)\, \dot{x}^2 & +& \nonumber \\ 
 2 \Lambda \, \dot{x}  & +& \nonumber  \\ 
 k \left( 1+ k_1 \, x + k_2 \, x^2\right) \!\!\! &  x &  \nonumber \\ 
 & = & \!\!  F(t).  \label{nonlin3}
\end{eqnarray}
In order to keep the writing lighter, the tilde is omitted in the core of the paper, leading finally to Eq. (\ref{dyneq}), Section \ref{exp}. Note however from Eq. (\ref{mis}) that a {\it heavy load} $m_l$ reduces the non-linear inertia terms $\tilde{m}_i$, while leaving the spring terms $k_i$ unchanged: a heavily loaded cantilever is thus very close to a Duffing oscillator. However, if the mass load is {\it zero}, the resonator {\it is not} a Duffing oscillator.


\end{appendix}

\end{document}